\documentclass[twocolumn]{aastex63}

\usepackage{CJKutf8}
\usepackage{amsmath}

\received{Aug 10, 2020}
\revised{Sep 12, 2020}
\accepted{Dec 2, 2020}
\shorttitle{Selected CT AGN in CDFS}
\shortauthors{Guo et al.}

\begin{document}

\title{Multi-wavelength Selected Compton-thick AGNs in \textit{Chandra} Deep Field-South Survey}

\correspondingauthor{Qiusheng Gu}
\email{qsgu@nju.edu.cn}

\author[0000-0002-2338-7709]{Xiaotong Guo \begin{CJK*}{UTF8}{gkai}(郭晓通)\end{CJK*}}
\affiliation{School of Astronomy and Space Science, Nanjing University, Nanjing, Jiangsu 210093, China}
\affiliation{Key Laboratory of Modern Astronomy and Astrophysics (Nanjing University), Ministry of Education, Nanjing 210093, China}

\author{Qiusheng Gu \begin{CJK*}{UTF8}{gkai}(顾秋生)\end{CJK*}}
\affiliation{School of Astronomy and Space Science, Nanjing University, Nanjing, Jiangsu 210093, China}
\affiliation{Key Laboratory of Modern Astronomy and Astrophysics (Nanjing University), Ministry of Education, Nanjing 210093, China}

\author{Nan Ding \begin{CJK*}{UTF8}{gkai}(丁楠)\end{CJK*}}
\affiliation{School of Physical Science and Technology, Kunming University, Kunming 650214, China}

\author{Xiaoling Yu \begin{CJK*}{UTF8}{gkai}(俞效龄)\end{CJK*}}
\affiliation{School of Astronomy and Space Science, Nanjing University, Nanjing, Jiangsu 210093, China}
\affiliation{Key Laboratory of Modern Astronomy and Astrophysics (Nanjing University), Ministry of Education, Nanjing 210093, China}

\author{Yongyun Chen \begin{CJK*}{UTF8}{gkai}(陈永云)\end{CJK*}}
\affiliation{College of Physics and Electronic Engineering, Qujing Normal University, Qujing 655011, China}


%
%
%
%

\nocollaboration{5}



\begin{abstract}
Even in deep X-ray surveys, Compton-thick active galactic nuclei (CT AGNs, ${\rm N_H} \geqslant 1.5~\times~10^{24}~{\rm cm}^{-2}$) are difficult to be identified due to X-ray flux suppression and their complex spectral shape. However, the study of CT AGNs is vital for understanding the rapid growth of black holes and the origin of cosmic X-ray background. In the local universe, the fraction of CT AGNs accounts for 30\% of the whole AGN population. We may expect a higher fraction of CT AGNs in deep X-ray surveys, however, only 10\% of AGNs have been identified as CT AGNs in the 7 Ms \textit{Chandra} Deep Field-South (CDFS) survey. In this work, we select 51 AGNs with abundant multi-wavelength data. 
Using the method of the mid-infrared (mid-IR) excess, we select hitherto unknown 8 CT AGN candidates in our sample. Seven of these candidates can confirm as CT AGN based on the multi-wavelength identification approach, and a new CT AGN (XID 133) is identified through the mid-IR diagnostics. We also discuss the X-ray origin of these eight CT AGNs and the reason why their column densities were underestimated in previous studies. We find that the multi-wavelength approaches of selecting CT AGNs are highly efficient, provided the high quality of observational data. We also find that CT AGNs have a higher Eddington ratio than non-CT AGNs, and that both CT AGNs and non-CT AGNs show similar properties of host galaxies.
\end{abstract}



\section{Introduction} \label{sec:intro}
It is well known that the X-ray emission has not only a strong pierce but also exhibits significant property of the particle.
The X-ray photon is absorbed as it passes through the interstellar medium due to Compton scattering.  
If the X-ray obscuring matter has an optical depth which is equal to or larger than 1 ($\tau = \sigma_{\rm T} \cdot {\rm N_H}$\footnote{$\sigma_{\rm T} $ is Thomson scattering cross-section ($\sigma_{\rm T} \approx 6.67\times10^{-25}~{\rm cm}^2$), ${\rm N_H}$ is the equivalent neutral hydrogen column density.} $ \geqslant 1$), then the active galactic nucleus (AGN) is called, by definition, Compton-thick (CT) AGN \citep[ i.e., $N_H \geqslant 1.5\times 10^{24}~{\rm cm}^{-2}$,][]{2004ASSL..308..245C}. 
The main characteristics of the X-ray spectra of CT AGNs are : (1) a flat spectrum with photon index, $\Gamma$, less than 1 at energies below 10 keV; (2) an absorption turnoff at energies above 10 keV, with the exact cut-off energy depending on the column density; and (3) a prominent iron K$\alpha$ emission line.

CT AGNs are believed to be in a phase of the evolutionary scenario of AGNs, which is a rapid growth state of central supermassive black holes \citep[SMBHs, e.g.,][]{2011MNRAS.411.1231G}.  
Some studies suggest a possible connection between CT AGNs and their host environments \citep[e.g.,][]{2017MNRAS.468.1273R}. For instance, CT AGNs prefer to host in gas-rich environments and galaxy mergers \citep{2015ApJ...814..104K}. Compared with Compton-thin AGNs, the host galaxies of CT AGNs show higher star formation rates \citep[SFRs, ][]{2012ApJ...755....5G}. Moreover, the cosmic X-ray background synthesis models \citep[e.g.,][]{2007A&A...463...79G,2012A&A...546A..98A} also require a significant fraction of CT AGNs. In the local universe, the fraction of CT AGNs is about 30\% of the total AGN population \citep[e.g.,][]{2015ApJ...815L..13R}. 
A higher fraction of CT AGNs is expected at high redshift, where the fraction of mass of atomic and/or molecular gas in galaxies is much higher than in the local universe \citep[e.g.,][]{2013ARA&A..51..105C}. However, CT AGNs are very difficult to be identified at high redshift, resulting in only a small fraction ($\lesssim$ 10\%) of CT AGNs is identified in deep X-ray surveys \citep[e.g.,][]{2016ApJ...817...34M, 2018MNRAS.480.2578L, 2015A&A...573A.137L}. Even Nuclear Spectroscopic Telescope Array \citep[NuSTAR,][]{2013ApJ...770..103H}, which is sensitive above 10 keV, was only able to increase the fraction of CT AGNs to $\sim 11.5\%$ in the UKIDSS Ultra Deep Survey field \citep{2018ApJS..235...17M}. Therefore, there should exist many CT AGNs that are still missed in deep X-ray surveys.

One important thing is how to identify CT AGNs efficiently. Currently, to identify CT AGN, the most commonly used method is X-ray spectroscopy fitting. For instance, 
\cite{2018MNRAS.480.2578L} selected 67 CT AGNs from 1855 point sources in the \textit{Chandra}-COSMOS. 
However, the major weakness of this method is the low efficiency of selecting CT AGN, and it depends relatively on the absorption-corrected model of X-ray.  
Thanks to the successful operation of NuSTAR, which can provide the X-ray source with a spectrum above 10~keV. The combination of NuSTAR and Chandra or XMM-Newton observations will constitute a broad waveband X-ray spectrum, which can be more effectively used to identify CT AGN.
Many recent studies have used the method of broad waveband X-ray spectroscopy fitting to identify CT AGNs \citep[e.g.,][]{2019ApJ...877..102K, 2019MNRAS.487.1662L, 2019ApJ...887..173L, 2020ApJ...888....8T}. 
Besides, several multi-wavelength techniques have also been developed to pre-select/identify CT AGNs in the past decade, based on the known CT AGN multi-band properties. The method of the mid-infrared (mid-IR) excess, among the multi-wavelength techniques, is usually used to select CT AGN candidates in some studies \citep[e.g.,][]{2011ApJ...740...37L, 2015A&A...578A.120L, 2019MNRAS.487.1662L}. The multi-wavelength techniques usually use the ratio between X-ray and other bands to identify CT AGNs, including X-ray to mid-IR luminosity ratio \citep[e.g.,][]{2015A&A...578A.120L, 2015ApJ...809..115L, 2017ApJ...846...20L} and X-ray to high-ionization optical emission lines (such as [O~III] and [Ne~V]) luminosity ratio \citep[e.g.,][]{1998A&A...338..781M, 2006A&A...446..459C, 2010A&A...519A..92G, 2015A&A...578A.120L}.

The \textit{Chandra} Deep Field-South (CDFS) survey is the deepest X-ray survey with an exposure time of about 7~Ms so far \citep{2017ApJS..228....2L}. There are 1008 sources detected by X-ray, among of which, 711 X-ray sources were classified as AGNs. 
The X-ray sources in the CDFS survey contain abundant photometric data from ultraviolet (UV) to infrared (IR), which were collected by \cite{2016ApJ...830...51S}. Some X-ray sources also contain optical spectra \citep[e.g.,][]{2004ApJS..155..271S, 2005A&A...437..883M, 2017A&A...608A...2I, 2017A&A...606A..12H} and radio data \citep[e.g.,][]{2008ApJS..179...71K, 2008ApJS..179..114M, 2013ApJS..205...13M}.
Since the CDFS survey is currently the most sensitive X-ray deep field, we expect that the fraction of CT AGNs can achieve as high as 30\% in this survey. However, only 71 CT AGNs (10\%) are identified by the method of X-ray spectroscopy fitting \citep{2017ApJS..232....8L, 2019ApJ...877....5L, 2019A&A...629A.133C}. Therefore, many CT AGNs ( about 20\%) have been missed by their methods in the CDFS survey. \cite{2020ApJ...897..160L} pointed out that a large population of obscured AGNs was mis-diagnosed as low-luminosity AGNs in the CDFS survey.
The motivation of our work is to find out missed CT AGNs in the CDFS survey. Moreover, we also try to understand why these CT AGNs are missed by X-ray spectroscopy fitting.

The structure of the paper is as follows. In Section~\ref{sec:sample}, we describe the sample and data. 
In Section~\ref{sec:select}, we pre-select 8 CT AGNs via the method of the mid-IR excess. 
The eight candidates will be identified by a combination of multi-wavelength identification approaches in Section~\ref{sec:diagno}.
In Section~\ref{sec:discussion}, we also discuss the X-ray origin of these eight CT AGNs, the reason why their column densities were underestimated, the efficiency of the multi-wavelength approaches of selecting CT AGNs, and comparing the properties of the CT AGNs and the non-CT AGNs.
Finally, we present a brief summary of this work in Section~\ref{sec:summary}.
We adopt a concordance flat $\Lambda$-cosmology with ${\rm H_0 = 67.4 \ km\ s^{-1}\ Mpc^{-1}}$, $\Omega_{\rm m} = 0.315$, and $\Omega_\Lambda = 0.685$ \citep{2020A&A...641A...6P}.
\section{The Sample and Data} \label{sec:sample}
\subsection{The AGN sample}

The starting point of the analysis is the X-ray sources in the CDFS survey. \cite{2016ApJ...830...51S} provided photometric catalogs (from UV to IR; a total of 43 bands) for the CDFS survey. In order to obtain the photometric data of the X-ray sources in the CDFS survey,
\cite{2020MNRAS.492.1887G} have cross-matched the X-ray catalog \citep{2017ApJS..228....2L} and the photometric catalog \citep{2016ApJ...830...51S} with a matching radius of 1 arcsec. The matching results contained 839 X-ray sources.
Based on the sample used in \cite{2020MNRAS.492.1887G}, we further select our AGN sample using the following three criteria:
\begin{itemize}
	\setlength{\itemsep}{0pt}
	\setlength{\parsep}{0pt}
	\setlength{\parskip}{0pt}
	\item[1)] A source is classified as AGN by X-ray. This source can also be selected as AGN by its spectral energy distribution \citep[SED,][]{2020MNRAS.495.1853P}. That is  to say, the confidence level of its AGN component, which is obtained by SED decomposition, is higher than 0.95.
	\item[2)] The photometric data of the source needs more than 20 bands, and mid to far-IR band photometry is required (IRAC, Spitzer/MIPS 24~\micron, Herschel/PACS 100~\micron, and 160~\micron), especially.
	\item[3)] We exclude merger sources identified by their optical images.
\end{itemize} 
The first criterion ensures that each source is an AGN. 
The second criterion guarantees that each source has enough photometric data to run the SED fitting model reliably.
The point spread function in mid-IR and far-IR is large. The merger sources are near each other. Thus, the flux density of the merger sources may be overestimated at mid-IR and far-IR bands. 
To be accurate photometric data, we implement the third criterion.
A final AGN sample (51 AGNs) is constructed, including eight known CT AGNs (XID 284, 332, 355, 405, 419, 587, 666, and 739), which are identified by the method of X-ray spectroscopy fitting \citep{2017ApJS..232....8L, 2019ApJ...877....5L, 2019A&A...629A.133C}.
\subsection{X-ray data}
The X-ray emission of the radio-quiet AGN is believed to arise from the corona above the accretion disk, and it is predominantly produced by the inverse Compton scattering of photons from the accretion disk.
Since the X-ray has strong pierced, the absorption-corrected X-ray luminosities with different correcting models for most AGNs are consistent. 
However, for CT AGN, there is a significant difference with differently correcting models \citep[e.g.,][]{2018A&A...618A..28Z}.
In some studies \citep[e.g.,][]{2017ApJS..228....2L}, the corrected luminosities are used to represent the intrinsic X-ray luminosities of AGNs.

Table~\ref{table:sample} shows our sample and the data used in this work. Column 5 of Table 1 lists the uncorrected 2--10 keV luminosities, which are converted by apparent 0.5--7 keV luminosities from \cite{2017ApJS..228....2L}\footnote{The 0.5--7 keV luminosities convert to 2--10 keV luminosities: ${\rm L}_{2\textendash 10}= {\rm L}_{0.5\textendash 7}({\rm ln} 10 - {\rm ln} 2)/({\rm ln} 7 - {\rm ln} 0.5), \Gamma = 2$; ${\rm L}_{2\textendash 10}= {\rm L}_{0.5\textendash 7}(10^{-\Gamma+2} - 2^{-\Gamma+2})/(7^{-\Gamma+2} - 0.5^{-\Gamma+2}), \Gamma \neq 2$. Where $\Gamma$ is photon index.}, where they estimated absorption by assuming that the intrinsic power-law spectrum had a fixed photon index of 1.8. For sources with effective photon indices greater than 1.8, the absorption column densities were set to zero. Using the estimated column densities, \cite{2017ApJS..228....2L} computed the absorption-corrected 0.5--7.0 keV luminosities. The absorption-corrected 0.5--7.0 keV luminosities also convert to 2--10 keV luminosities listed in Column 6 of Table~\ref{table:sample}. \cite{2017ApJS..232....8L} performed an X-ray spectral analysis for the bright sources  in the CDFS using the \textit{wabs * (zwabs*powerlaw + zgauss + powerlaw + zwabs*pexrav*constant)} model, and provide the absorption-corrected 2--10 keV luminosities listed in Column 7 of Table~\ref{table:sample}. To search for a best fit, \cite{2019ApJ...877....5L} adopted two models to fit the source spectra. Column 8 of Table~\ref{table:sample} lists the absorption-corrected 2--10 keV luminosities provided by \cite{2019ApJ...877....5L}. \cite{2019A&A...629A.133C} selected 20 CT AGNs using the automated spectral analysis. Column 9 of Table~\ref{table:sample} lists the absorption-corrected 2--10 keV luminosities of 3 CT AGNs, which are obtained by \cite{2019A&A...629A.133C}.

Moreover, this work also uses the X-ray spectra of the eight CT AGN candidates. Their X-ray spectra are the merged spectra for which the 102 observations are matched to an identically astrometric frame. In this work, we use the X-ray spectra produced by \cite{2017ApJS..228....2L}. For more detailed data reduction, please see Section 2.2 of \cite{2017ApJS..228....2L}.

\begin{deluxetable*}{l l l l l l l l l l l l l}
		\tablecaption{Physical properties of our AGN sample.\label{table:sample}}
		\tablewidth{700pt}
		\tabletypesize{\footnotesize}
		\tablehead{
			\colhead{XID} & \colhead{RA} & \colhead{DEC} & \colhead{z} &
			\colhead{L$_{\rm X}$\_App} & \colhead{L$_{\rm X}$\_Luo} & 
			\colhead{L$_{\rm X}$\_Liu} & \colhead{L$_{\rm X}$\_Li} & 
			\colhead{L$_{\rm X}$\_Cor} & \colhead{$\nu$L$_\nu$(6~\micron)} & \colhead{M$_{\rm BH}$} & \colhead{L$_{\rm bol}$} & \colhead{$\lambda_{\rm Edd}$} \\ 
			\colhead{} &  \colhead{[degree]} & \colhead{[degree]} &\colhead{} & \colhead{[erg~s$^{-1}$]} & \colhead{[erg~s$^{-1}$]}  & \colhead{[erg~s$^{-1}$]} & \colhead{[erg~s$^{-1}$]} & \colhead{[erg~s$^{-1}$]} & \colhead{[erg~s$^{-1}$]} & \colhead{[M$_\odot$]} & \colhead{[erg~s$^{-1}$]} & \colhead{}\\
			\colhead{(1)} & \colhead{(2)} & \colhead{(3)} & \colhead{(4)}  & \colhead{(5)} & \colhead{(6)} & \colhead{(7)} & \colhead{(8)} & \colhead{(9)} & \colhead{(10)} & \colhead{(11)} & \colhead{(12)} & \colhead{(13)}
		} 
		\startdata
		100 & 53.006054 & -27.694009 & 1.41 & 7.427e+43 & 1.153e+44 & 1.047e+44 & \multicolumn{1}{c}{\nodata} & \multicolumn{1}{c}{\nodata} & 2.200e+44 & 1.226e+08 & 1.471e+45 & 0.092\\
		101 & 53.006491 & -27.734132 & 3.37 & 9.177e+42 & 7.866e+43 & 7.413e+43 & 7.413e+43 & \multicolumn{1}{c}{\nodata} & 9.475e+44 & 9.918e+07 & 5.079e+45 & 0.394\\
		119 & 53.015280 & -27.767692 & 0.57 & 9.310e+42 & 1.748e+43 & 1.230e+43 & \multicolumn{1}{c}{\nodata} & \multicolumn{1}{c}{\nodata} & 1.685e+43 & 2.448e+08 & 8.338e+43 & 0.003\\
		133 & 53.020614 & -27.742022 & 3.47 & 1.463e+42 & 6.200e+43 & \multicolumn{1}{c}{\nodata} & 8.913e+43 & \multicolumn{1}{c}{\nodata} & 3.288e+44 & 1.565e+07 & 4.914e+45 & 2.415\\
		135 & 53.022332 & -27.778865 & 2.52 & 1.830e+43 & 1.622e+44 & 1.635e+44 & 1.107e+45 & \multicolumn{1}{c}{\nodata} & 8.647e+44 & 3.424e+08 & 4.265e+45 & 0.096\\
		167 & 53.031730 & -27.870385 & 0.81 & 4.663e+42 & 6.795e+42 & 5.495e+42 & \multicolumn{1}{c}{\nodata} & \multicolumn{1}{c}{\nodata} & 3.612e+43 & 6.266e+07 & 2.441e+44 & 0.030\\
		169 & 53.033222 & -27.710843 & 0.55 & 1.820e+42 & 2.394e+42 & 3.890e+42 & \multicolumn{1}{c}{\nodata} & \multicolumn{1}{c}{\nodata} & 3.450e+43 & 4.801e+07 & 2.477e+44 & 0.040\\
		175 & 53.036121 & -27.792852 & 0.54 & 7.993e+43 & 8.058e+43 & 7.762e+43 & \multicolumn{1}{c}{\nodata} & \multicolumn{1}{c}{\nodata} & 4.261e+44 & 2.627e+08$^*$ & 6.368e+45 & 0.186\\
		200 & 53.043796 & -27.719170 & 0.30 & 3.244e+42 & 3.301e+42 & 3.227e+42 & \multicolumn{1}{c}{\nodata} & \multicolumn{1}{c}{\nodata} & 8.077e+42 & 1.037e+07 & 6.774e+43 & 0.050\\
		214 & 53.047424 & -27.870404 & 3.74 & 2.591e+44 & 6.934e+44 & 8.128e+44 & 8.913e+44 & \multicolumn{1}{c}{\nodata} & 1.737e+45 & 5.967e+08 & 5.727e+45 & 0.074\\
		221 & 53.049089 & -27.774479 & 1.60 & 2.341e+43 & 3.407e+43 & 3.631e+43 & \multicolumn{1}{c}{\nodata} & \multicolumn{1}{c}{\nodata} & 1.408e+44 & 5.436e+07 & 8.903e+44 & 0.126\\
		249 & 53.057811 & -27.713630 & 0.73 & 5.855e+42 & 2.371e+43 & 1.950e+43 & 1.288e+43 & \multicolumn{1}{c}{\nodata} & 4.768e+43 & 1.710e+08 & 3.220e+44 & 0.014\\
		252 & 53.058359 & -27.850128 & 0.12 & 5.675e+41 & 5.531e+41 & 4.467e+41 & \multicolumn{1}{c}{\nodata} & \multicolumn{1}{c}{\nodata} & 2.658e+41 & 2.824e+06 & 2.979e+42 & 0.008\\
		284 & 53.065940 & -27.701854 & 2.07 & 2.744e+42 & 4.394e+43 & 1.862e+44 & 2.818e+44 & \multicolumn{1}{c}{\nodata} & 6.020e+44 & 1.678e+08 & 2.690e+45 & 0.123\\
		332 & 53.076835 & -27.765534 & 0.73 & 4.898e+41 & 1.695e+42 & 1.148e+43 & 2.399e+43 & 1.585e+43 & 3.569e+43 & 1.942e+08 & 2.816e+44 & 0.011\\
		341 & 53.079438 & -27.741622 & 1.83 & 1.098e+42 & 3.324e+42 & \multicolumn{1}{c}{\nodata} & \multicolumn{1}{c}{\nodata} & \multicolumn{1}{c}{\nodata} & 1.768e+45 & 1.521e+08 & 1.007e+46 & 0.509\\
		342 & 53.079505 & -27.870571 & 0.76 & 7.910e+40 & 7.910e+40 & \multicolumn{1}{c}{\nodata} & \multicolumn{1}{c}{\nodata} & \multicolumn{1}{c}{\nodata} & 2.423e+43 & 8.154e+06 & 1.533e+44 & 0.145\\
		355 & 53.082573 & -27.689614 & 0.23 & 1.495e+41 & 1.591e+41 & 8.511e+47 & 1.202e+43 & \multicolumn{1}{c}{\nodata} & 1.540e+43 & 6.302e+07 & 1.215e+44 & 0.015\\
		358 & 53.083579 & -27.746440 & 1.91 & 9.296e+42 & 1.484e+44 & 1.778e+44 & 2.291e+44 & \multicolumn{1}{c}{\nodata} & 1.183e+45 & 1.198e+08 & 7.897e+45 & 0.507\\
		399 & 53.093942 & -27.767736 & 1.73 & 1.969e+43 & 9.274e+43 & 6.457e+43 & 4.365e+43 & \multicolumn{1}{c}{\nodata} & 3.033e+44 & 1.002e+08 & 1.221e+45 & 0.094\\
		405 & 53.094921 & -27.757726 & 1.99 & 6.134e+41 & 7.798e+42 & \multicolumn{1}{c}{\nodata} & 1.479e+44 & \multicolumn{1}{c}{\nodata} & 1.144e+44 & 6.352e+07 & 9.030e+44 & 0.109\\
		419 & 53.097664 & -27.715210 & 2.15 & 2.196e+42 & 3.084e+43 & 3.631e+43 & 3.020e+44 & \multicolumn{1}{c}{\nodata} & 1.026e+45 & 2.105e+08 & 7.232e+45 & 0.264\\
		428 & 53.101059 & -27.690714 & 0.53 & 3.868e+42 & 4.077e+42 & 3.981e+42 & \multicolumn{1}{c}{\nodata} & \multicolumn{1}{c}{\nodata} & 4.766e+42 & 1.733e+08 & 6.183e+43 & 0.003\\
		433 & 53.102598 & -27.860506 & 0.70 & 4.196e+40 & 4.196e+40 & \multicolumn{1}{c}{\nodata} & \multicolumn{1}{c}{\nodata} & \multicolumn{1}{c}{\nodata} & 4.355e+42 & 2.207e+05 & 2.482e+43 & 0.865\\
		458 & 53.107040 & -27.718233 & 2.29 & 8.103e+43 & 1.909e+44 & 1.622e+44 & 1.862e+44 & \multicolumn{1}{c}{\nodata} & 3.058e+45 & 3.820e+08 & 1.343e+46 & 0.270\\
		479 & 53.110406 & -27.676587 & 1.03 & 9.784e+43 & 9.784e+43 & 9.550e+43 & \multicolumn{1}{c}{\nodata} & \multicolumn{1}{c}{\nodata} & 8.867e+44 & 2.738e+08$^*$ & 1.274e+46 & 0.358\\
		488 & 53.111539 & -27.695956 & 0.73 & 5.508e+42 & 7.929e+42 & 8.913e+42 & \multicolumn{1}{c}{\nodata} & \multicolumn{1}{c}{\nodata} & 7.288e+43 & 1.460e+08 & 4.441e+44 & 0.023\\
		507 & 53.115101 & -27.695854 & 0.67 & 1.600e+43 & 2.708e+43 & 2.399e+43 & \multicolumn{1}{c}{\nodata} & \multicolumn{1}{c}{\nodata} & 6.815e+43 & 8.333e+07 & 4.919e+44 & 0.045\\
		546 & 53.123843 & -27.862651 & 1.42 & 4.148e+42 & 3.252e+43 & 2.818e+43 & 5.495e+43 & \multicolumn{1}{c}{\nodata} & 4.603e+43 & 1.562e+07 & 2.666e+44 & 0.131\\
		549 & 53.124083 & -27.891171 & 2.71 & 4.517e+42 & 7.496e+43 & 6.576e+43 & 1.117e+44 & \multicolumn{1}{c}{\nodata} & 3.258e+44 & 2.614e+08 & 2.175e+45 & 0.064\\
		557 & 53.124917 & -27.758309 & 1.21 & 6.556e+43 & 8.035e+43 & 7.762e+43 & \multicolumn{1}{c}{\nodata} & \multicolumn{1}{c}{\nodata} & 3.491e+44 & 2.566e+09$^*$ & 6.400e+45 & 0.019\\
		567 & 53.125902 & -27.751285 & 0.74 & 2.429e+43 & 2.429e+43 & 2.455e+43 & \multicolumn{1}{c}{\nodata} & \multicolumn{1}{c}{\nodata} & 2.541e+43 & 5.304e+08$^*$ & 6.356e+44 & 0.009\\
		587 & 53.131130 & -27.773118 & 2.22 & 3.559e+42 & 9.948e+42 & \multicolumn{1}{c}{\nodata} & \multicolumn{1}{c}{\nodata} & 7.943e+43 & 1.567e+45 & 4.561e+08 & 9.921e+45 & 0.167\\
		623 & 53.139317 & -27.874410 & 3.88 & 7.486e+42 & 2.237e+43 & \multicolumn{1}{c}{\nodata} & 3.890e+43 & \multicolumn{1}{c}{\nodata} & 9.523e+45 & 3.079e+08 & 4.887e+46 & 1.221\\
		640 & 53.142749 & -27.827834 & 0.55 & 2.952e+40 & 2.952e+40 & \multicolumn{1}{c}{\nodata} & \multicolumn{1}{c}{\nodata} & \multicolumn{1}{c}{\nodata} & 3.972e+42 & 2.312e+05 & 2.676e+43 & 0.890\\
		652 & 53.145607 & -27.919722 & 0.84 & 1.116e+43 & 1.267e+43 & 1.288e+43 & \multicolumn{1}{c}{\nodata} & \multicolumn{1}{c}{\nodata} & 3.481e+43 & 6.237e+07 & 2.681e+44 & 0.033\\
		666 & 53.148800 & -27.821098 & 2.58 & 5.196e+42 & 2.481e+43 & 4.898e+43 & 4.169e+44 & 3.981e+44 & 4.554e+45 & 3.724e+08 & 2.419e+46 & 0.500\\
		691 & 53.152951 & -27.735139 & 0.67 & 9.102e+42 & 9.985e+42 & 9.120e+42 & \multicolumn{1}{c}{\nodata} & \multicolumn{1}{c}{\nodata} & 1.332e+43 & 6.673e+07 & 8.171e+43 & 0.009\\
		711 & 53.157312 & -27.870054 & 1.60 & 1.735e+44 & 1.805e+44 & 2.239e+44 & \multicolumn{1}{c}{\nodata} & \multicolumn{1}{c}{\nodata} & 1.687e+45 & 7.801e+08 & 9.794e+45 & 0.097\\
		715 & 53.158443 & -27.773968 & 2.12 & 3.569e+43 & 3.888e+43 & \multicolumn{1}{c}{\nodata} & \multicolumn{1}{c}{\nodata} & \multicolumn{1}{c}{\nodata} & 4.314e+44 & 1.663e+08 & 1.509e+45 & 0.070\\
		716 & 53.158811 & -27.662418 & 0.84 & 8.554e+43 & 8.554e+43 & 9.550e+43 & \multicolumn{1}{c}{\nodata} & \multicolumn{1}{c}{\nodata} & 2.165e+44 & 5.276e+08$^*$ & 3.878e+45 & 0.057\\
		735 & 53.162839 & -27.767166 & 1.22 & 3.449e+43 & 4.792e+43 & 3.715e+43 & \multicolumn{1}{c}{\nodata} & \multicolumn{1}{c}{\nodata} & 2.468e+44 & 1.957e+09$^*$ & 6.027e+45 & 0.024\\
		739 & 53.163574 & -27.890375 & 2.28 & 2.362e+42 & 2.040e+43 & \multicolumn{1}{c}{\nodata} & 1.353e+44 & \multicolumn{1}{c}{\nodata} & 2.464e+45 & 2.378e+08 & 1.553e+46 & 0.502\\
		746 & 53.165266 & -27.814067 & 3.06 & 8.921e+43 & 5.800e+44 & 4.169e+44 & 6.457e+44 & \multicolumn{1}{c}{\nodata} & 1.843e+45 & 3.985e+08 & 8.560e+45 & 0.165\\
		760 & 53.170107 & -27.929669 & 3.35 & 9.286e+43 & 8.055e+44 & 5.754e+44 & 5.248e+44 & \multicolumn{1}{c}{\nodata} & 8.943e+45 & 6.781e+08 & 5.028e+46 & 0.570\\
		802 & 53.181858 & -27.814094 & 2.41 & 8.342e+41 & 5.446e+42 & \multicolumn{1}{c}{\nodata} & 1.047e+43 & \multicolumn{1}{c}{\nodata} & 2.235e+45 & 7.714e+07 & 1.273e+46 & 1.269\\
		844 & 53.195120 & -27.855704 & 1.10 & 7.647e+41 & 1.110e+42 & \multicolumn{1}{c}{\nodata} & 1.577e+42 & \multicolumn{1}{c}{\nodata} & 6.150e+43 & 2.228e+07 & 4.143e+44 & 0.143\\
		855 & 53.199525 & -27.709118 & 0.98 & 1.060e+44 & 1.379e+44 & 1.318e+44 & \multicolumn{1}{c}{\nodata} & \multicolumn{1}{c}{\nodata} & 4.623e+44 & 5.313e+08 & 3.484e+45 & 0.050\\
		867 & 53.205002 & -27.680657 & 1.22 & 1.216e+43 & 3.858e+43 & 6.026e+43 & 1.820e+44 & \multicolumn{1}{c}{\nodata} & 2.352e+45 & 5.406e+08 & 1.187e+46 & 0.169\\
		889 & 53.216307 & -27.743335 & 0.52 & 1.922e+41 & 1.951e+41 & \multicolumn{1}{c}{\nodata} & \multicolumn{1}{c}{\nodata} & \multicolumn{1}{c}{\nodata} & 1.024e+44 & 1.609e+08 & 6.915e+44 & 0.033\\
		933 & 53.247142 & -27.816305 & 1.65 & 3.328e+43 & 1.645e+44 & 1.349e+44 & 1.288e+44 & \multicolumn{1}{c}{\nodata} & 2.702e+45 & 4.962e+08 & 1.338e+46 & 0.207\\
		\enddata
		\tablecomments{(1) is X-ray ID from the \cite{2017ApJS..228....2L} catalog. (2) and (3) are the R.A. and decl. of the source, respectively. (4) is redshift from \cite{2020MNRAS.492.1887G}. 
			(5) and (6) are the apparent rest-frame 2--10 keV luminosity and absorption-corrected 2--10 keV luminosity that is converted by apparent and absorption-corrected 0.5--7 keV luminosity from \cite{2017ApJS..228....2L}. 
			(7), (8) and (9) are the absorption-corrected 2--10 keV luminosity by \cite{2017ApJS..232....8L}, \cite{2019ApJ...877....5L} and Corral et al. (2019), respectively. (10) is the rest-frame luminosity of the AGN component at 6~\micron. (11), (12) and 13 are BH mass, bolometric luminosity and Eddington ratio. The $^*$ represents the BH mass estimated with broad line.}
\end{deluxetable*}

\subsection{Multi-band SED fitting and Mid-IR data}

The photometric data from UV to IR can constitute a multi-band SED, which is composed of multiple thermal emission components. 
Thanks to SED fitting technique, we can perform a detailed multi-band SED decomposition, including the stellar radiation, the dust re-radiation, and AGN emission. 
SED fitting technique can also constrain some crucially physical quantities of both AGNs and host galaxies, such as AGN luminosity, AGN fraction, SFR, and stellar mass (M$_*$). 
\cite{2020MNRAS.492.1887G} obtained some physical quantities (including AGN luminosity, AGN fraction, SFR, M$_*$, dust luminosity, AGN type, and rest-frame 6~\micron\ luminosity for AGN component) of all AGNs in our sample through SED fitting using photometric data from UV to IR. Therefore, some physical quantities used in this work are from \cite{2020MNRAS.492.1887G}.

The mid-IR emission (3--30 \micron) of an AGN is most distinct from that of its host galaxy. 
AGNs are usually bright in the mid-IR waveband because of the thermal emission from hot dust in the torus, which is heated by absorbing UV and optical photons from the accretion disk. 
The optical depth is low at the mid-IR waveband, and therefore the mid-IR emission of the AGN is not strongly suppressed. 
In this paper, we use the 6~\micron\ emission to represent the mid-IR emission of the AGN \citep[e.g.,][]{2004A&A...418..465L, 2017ApJ...837..145C}, which is obtained by SED decomposition. Column 10 of Table~\ref{table:sample} lists the 6~\micron\ luminosities of AGNs, provided by \cite{2020MNRAS.492.1887G}.

\section{Mid-IR excess pre-selecting CT AGNs}\label{sec:select}
The mid-IR and X-ray luminosities of AGNs are closely related. Since the X-ray emission of CT AGNs is strongly suppressed, the intrinsic X-ray luminosities of CT AGNs are usually underestimated. The mid-IR emission of the AGNs is not suppressed because the optical depth is low. In order to select CT AGNs, one easy way is to find the AGNs whose intrinsic X-ray luminosities are underestimated. 
\subsection{The method of Mid-IR excess}\label{sec:method}

The mid-IR and X-ray emission of radio-quiet AGNs are both excellent tracers of supermassive black hole accretion power.
Previous studies have investigated the correlation between the mid-IR and X-ray emission \citep[e.g.,][]{2004A&A...418..465L, 2009A&A...498...67L, 2009A&A...502..457G, 2009ApJ...693..447F, 2017ApJ...837..145C}. For instance, \cite{2015ApJ...807..129S} derived a relation for the radio-quiet AGNs across a wide range of luminosity (from local Seyfert galaxies with L$_X\sim 10^{42}$ erg s$^{-1}$  to the most luminous quasars with L$_X\sim 10^{46}$ erg s$^{-1}$), which is 
\begin{equation}
	{\rm log\ L}(2\textendash10\ {\rm keV})= 40.981 + 1.024 x -0.047 x^2 ,
	\label{equ:relation}
\end{equation}
where L(2--10 keV) is in unit of erg~s$^{-1}$ and $x \equiv {\rm log} (\nu {\rm L}_{\nu} (6~\micron)/10^{41}$ erg~s$^{-1}$).
In \cite{2015ApJ...807..129S}, the unobscured and Compton-thin AGNs are in good agreement with this relation. The relation should also be appropriate for CT AGNs, while the X-ray luminosities of CT AGNs mentioned by \cite{2015ApJ...807..129S} are much lower than their mid-IR luminosities. The reason is that the absorption-corrected X-ray luminosities of CT AGNs are more model-dependent. Thus their intrinsic X-ray luminosities are underestimated. 
Indeed, the method of mid-IR excess is suitable to pre-select CT AGNs whose intrinsic X-ray luminosities are strongly underestimated.

The above mentioned are all radio-quiet AGNs. For the radio-loud AGNs, enhanced X-ray emission is usually observed in radio-loud AGNs \citep[e.g.,][]{2011ApJ...726...20M,2012A&A...545A..66B} due to the contribution of the jets to the total X-ray emission. While the mid-IR emission of radio-loud AGNs (excluding blazars) is similar to that of radio-quiet AGNs. The above method of mid-IR excess can be also used to select CT AGN candidates of radio-loud. 
There are nine radio-loud AGNs\footnote{Radio-loud AGNs: XID 119, 175, 342, 587, 623, 746, 760, 844, and 867.} in our sample, including 2 CT AGNs that are identified by this work.
The radio-loud AGNs will less impact on the result of mid-IR excess selection.


\subsection{CT AGN candidates}

\begin{figure*}[ht]
	\includegraphics[width=0.5\linewidth]{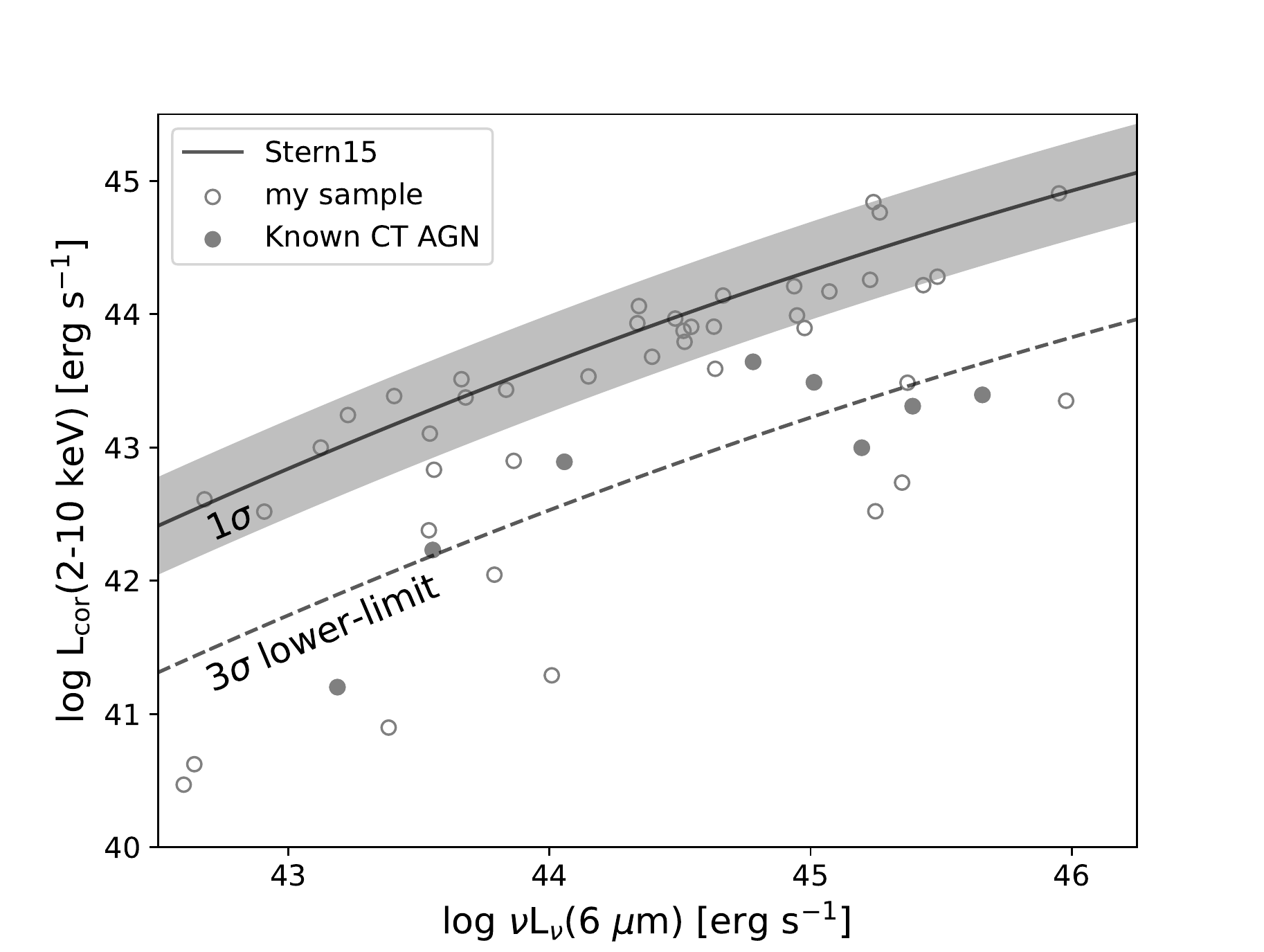}
	\includegraphics[width=0.5\linewidth]{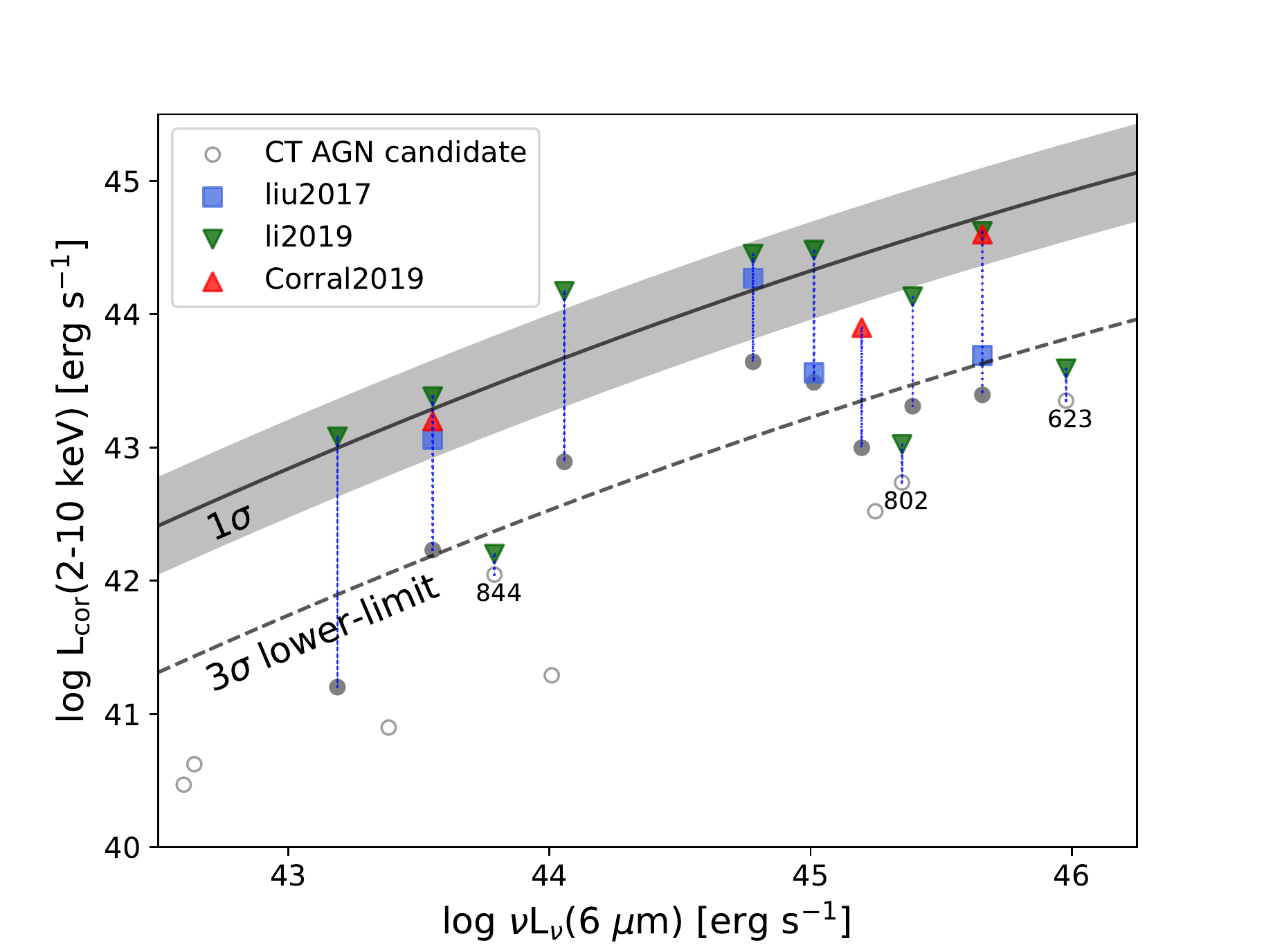}
	
	\caption{Left panel: Rest-frame 6~\micron\ versus absorption-corrected 2--10 keV luminosities for our AGN sample. The black line represents the relation for \cite{2015ApJ...807..129S}(Equation~\ref{equ:relation}). The gray shaded area is 1$\sigma$ dispersion, the dashed line is 3$\sigma$ lower limit. Open grey circles show our AGN sample, the solid grey circles represent the eight known CT AGNs. Right panel: Rest-frame 6~\micron\  versus different absorption-corrected 2--10 keV X-ray luminosities for 8 known CT AGNs and 8 CT AGN candidates. Grey circles, blue square, green inverted triangles and red triangles represent different X-ray luminosities corrected by \cite{2017ApJS..228....2L}, \cite{2017ApJS..232....8L}, \cite{2019ApJ...877....5L} and \cite{2019A&A...629A.133C}.}
	\label{fig:relation-selection}
\end{figure*}

The left panel of Figure~\ref{fig:relation-selection} shows that the mid-IR versus absorption-corrected X-ray luminosities. The absorption-corrected X-ray luminosities are estimated with a simple model \citep[see Section 4.4 in ][]{2017ApJS..228....2L}.
The black line represents the relation of Equation~\ref{equ:relation}. We obtain the 1$\sigma$ uncertainty of the \cite{2015ApJ...807..129S} relation with using the data in Tables of \cite{2015ApJ...807..129S}. The gray shaded area is 1$\sigma$ dispersion for the \cite{2015ApJ...807..129S} relation \citep[also see][]{2020MNRAS.499.1823Z}. 
The solid grey circles represent the eight known CT AGNs. The dashed line is 3$\sigma$ lower limit. 
Most of the AGNs locate near the black line with 1$\sigma$ area. 
However, there are still 12 AGNs below the 3$\sigma$ lower limit, four of which are the known CT AGNs.
The remaining 8 AGNs are also likely to be CT AGNs (hereafter referred to as CT AGN candidates).

The right panel of Figure~\ref{fig:relation-selection} shows that the rest-frame 6~\micron\ versus various absorption-corrected 2--10 keV luminosities of 8 known CT AGNs and 8 CT AGN candidates. 
Except for three known CT AGNs, the remaining five known CT AGNs corrected by \cite{2017ApJS..232....8L}, \cite{2019ApJ...877....5L}, or \cite{2019A&A...629A.133C} are near the black line with 1$\sigma$ area. The absorption-corrected luminosities of the three known CT AGNs are above 3$\sigma$ lower limit, suggesting that their absorption correction should be reasonable.
However, among 8 CT AGN candidates, only three candidates (XID 623, 802, and 844) are corrected by \cite{2019ApJ...877....5L}.
Their absorption-corrected luminosities are still below the dashed line, suggesting that their column densities may be underestimated.
There are still five candidates that are not corrected by previous works \citep{2017ApJS..232....8L,2019ApJ...877....5L,2019A&A...629A.133C}. 
To determine whether these eight candidates are CT AGNs, we will perform X-ray diagnostics, optical spectral diagnostics, and mid-IR diagnostics on these eight candidates.

\section{Multi-wavelength diagnostics}\label{sec:diagno}

\subsection{X-ray diagnostics}\label{sec:X-ray-diagno}
The X-ray diagnostics is the most reliable identification method for CT AGN \citep[e.g.,][]{2017AN....338..316L}.
\cite{2019ApJ...877....5L} have fitted the X-ray spectra of source XID 623, 802, and 884 using MYTorus, but did not confirm them as CT AGN. We believe that it is difficult to obtain reliable evidence through X-ray spectroscopy fitting to prove that these eight candidates are CT AGNs. 
Therefore, we only intend to use the characteristics of their X-ray spectra, in this section, to search for supporting evidence.

We fit the X-ray spectra of these eight candidates in Xspec v 12.11.0 \citep{1996ASPC..101...17A} using the Cash statistic that is more appropriate with the low count regime \citep{1979ApJ...228..939C}. Due to the limited counts of these eight candidates, we do not bin the spectrum because it might lose some information. 
In order to obtain the characteristics of the X-ray spectrum at low-energy (2--10 keV) and high-energy ($>$ 10 keV),
we use \textit{phabs * (powerlaw + zgauss)}  to fit the rest-frame 2--10 keV spectra, and use \textit{phabs * powerlaw} to fit the rest-frame above 10 keV spectra. The \textit{phabs} accounts for the Galactic absorption, which is fixed at a column density of $8.8 \times 10^{-19} \ {\rm cm}^{-2}$ \citep{1992ApJS...79...77S}. 
The characteristics of the X-ray spectra of these eight candidates are derived by fitting their spectra.
Their properties and the characteristics of their X-ray spectra are listed in Table~\ref{table:x-ray-diagno}.

\begin{deluxetable*}{l l c r c c r c}[ht]
	\tablecaption{The characteristics of X-ray spectra.\label{table:x-ray-diagno}}
	\tablewidth{700pt}
	\tabletypesize{\footnotesize}
	\tablehead{
		\colhead{XID} & \colhead{z} & 
		\colhead{${\rm L_{obs}}$(2--10 keV)}& 
		\colhead{$\Gamma$(2--10 keV)}&\colhead{$Fe\ K\alpha$}& \colhead{C-stat/d.o.f.(2--10 keV)} & \colhead{$\Gamma$($>$10 keV)} & 
		\colhead{C-stat/d.o.f.($>$10 keV)} \\ 
		\colhead{}&\colhead{}  & \colhead{[erg~s$^{-1}$]}&\colhead{} & \colhead{} & \colhead{}\\
		\colhead{(1)} & \colhead{(2)} & \colhead{(3)} & \colhead{(4)}  & \colhead{(5)} & \colhead{(6)} & \colhead{(7)} & \colhead{(8)}
	} 
	\startdata
	341 & 1.83 & $8.5^{+2.6}_{-2.8}\times 10^{41}$ & $0.72 \pm 0.81$ & no  & 139.7/189&$-0.12 \pm 1.69$ & 150.4/237  \\
	342 & 0.76 & $4.6^{+2.4}_{-2.4}\times 10^{40}$ &$5.34 \pm 2.79$ &no&227.7/312 & \multicolumn{1}{c}{\nodata} & \multicolumn{1}{c}{\nodata} \\
	433 & 0.70 & $1.4^{+0.1}_{-1.4}\times 10^{40}$ &$5.33 \pm 6.40$ &no&142.8/319 & \multicolumn{1}{c}{\nodata} & \multicolumn{1}{c}{\nodata} \\
	623 & 3.88 & $6.8^{+1.0}_{-1.2}\times 10^{42}$ &$-0.21 \pm 0.83$ &yes&92.3/99& $1.84 \pm 0.77$ & 237.1/340 \\
	640 & 0.54 &$2.8^{+0.9}_{-1.6}\times 10^{40}$ & $3.49 \pm 1.64$ &no&119.2/353 & \multicolumn{1}{c}{\nodata} & \multicolumn{1}{c}{\nodata} \\
	802 & 2.41 & $5.4^{+2.5}_{-4.2}\times 10^{41}$ &$6.07 \pm 1.92$ &no&89.7/161 & $1.35 \pm 1.35$ & 137.8/278 \\
	844 & 1.10 &$9.9^{+2.4}_{-2.7}\times 10^{41}$ & $0.80 \pm 0.36$ &no&244.3/264 & $0.73 \pm 6.56$ & 105.5/148 \\
	889 & 0.52 & $1.9^{+0.6}_{-0.7}\times 10^{41}$ & $1.53 \pm 0.53$ &no&372.4/360 & \multicolumn{1}{c}{\nodata} & \multicolumn{1}{c}{\nodata} \\
	\enddata
	\tablecomments{(1) and (2) are XID and redshift. (3) is the apparent rest-frame 2--10 keV luminosity with 1$\sigma$ confidence derived by fitting X-ray spectra. (4) is photon index of rest-frame 2--10 keV. (5) represents whether there is a characteristic of the prominent iron K$\alpha$ emission line in the X-ray spectrum. (6) is C-stat/d.o.f. of the best fit for rest-frame 2--10 keV. (7) is photon index above 10 keV. (8) is C-stat/d.o.f. of the best fit above 10 keV.}
\end{deluxetable*}

Three candidates (XID 341, 623, and 844) have a flat spectrum with $\Gamma < 1$ at energies below 10 keV. There are four candidates with X-ray spectra above 10 keV. However, only the X-ray spectrum of the source XID 623 is relatively reliable and shows a characteristic of the prominent iron K$\alpha$ emission line with the equivalent width of $0.234$ keV. Therefore, only the source XID 623 is more likely a CT AGN in terms of its X-ray spectrum characteristics.
	
\subsection{Optical spectral diagnostics}
The torus of an AGN does not obscure the high-ionization narrow optical emission lines from the narrow-line regions (NLRs). As long as the unified scheme holds, narrow emission lines have been considered to be a useful indicator of the AGN luminosity. Therefore, there should be a correlation between the narrow emission lines and X-ray luminosities, and the relationship has been investigated \citep[e.g.,][]{2005ApJ...634..161H}.
In this section, we use the high-ionization narrow optical emission lines to identify whether these eight candidates are CT AGNs. We try to collect the optical spectra of these eight candidates from the literature \citep[e.g.,][]{2004ApJS..155..271S, 2005A&A...437..883M, 2017A&A...608A...2I, 2017A&A...606A..12H}, and finally we only find two spectra of XID 640 and 889. The spectrum of the source XID 640 originates from \cite{2017A&A...606A..12H}, and XID 889 from Figure~6 (Source 175b) in \cite{2004ApJS..155..271S}. There is only [O III]$\lambda 5007$ emission line detected in the spectrum of source XID 640. We fit the [O III]$\lambda 5007$ emission line and obtain its flux, $f_{\rm [O~III]}=1.57\times 10^{-17}$ erg~s$^{-1}$~cm$^{-2}$. In addition, we estimate the flux of the high-ionization optical emission lines ([O~III]$\lambda 5007$ and [Ne~V]$\lambda 3426$) in the spectrum of source XID 889, i.e. $f_{\rm [O~III]}=8.93\times 10^{-16}$  erg~s$^{-1}$~cm$^{-2}$ and $f_{\rm [Ne~V]}=1.26\times 10^{-16}$  erg~s$^{-1}$~cm$^{-2}$, respectively. In this section, the observed X-ray luminosities are from Section~\ref{sec:X-ray-diagno}.

\begin{figure*}[ht]
	\includegraphics[width=0.5\linewidth]{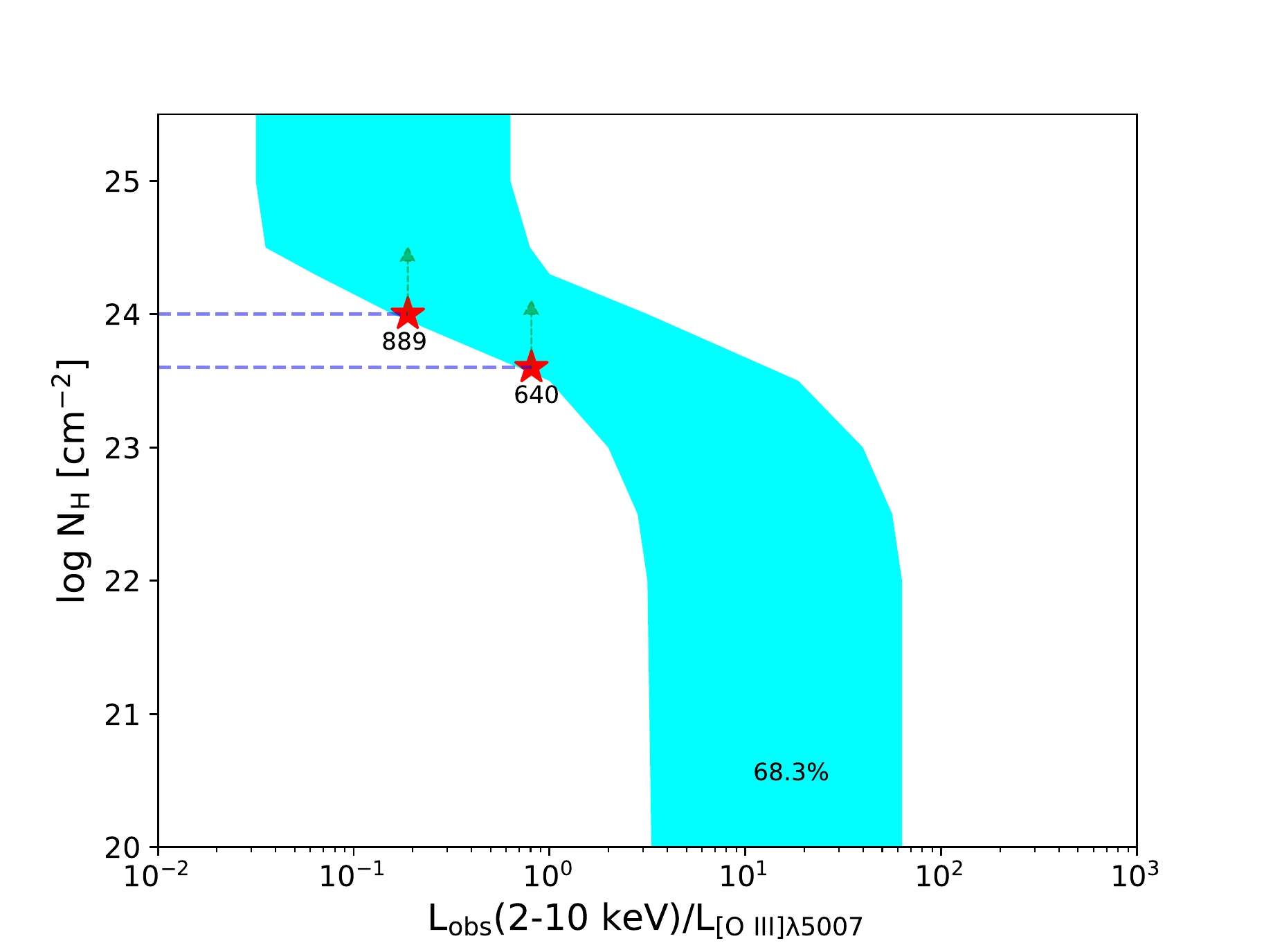}
	\includegraphics[width=0.5\linewidth]{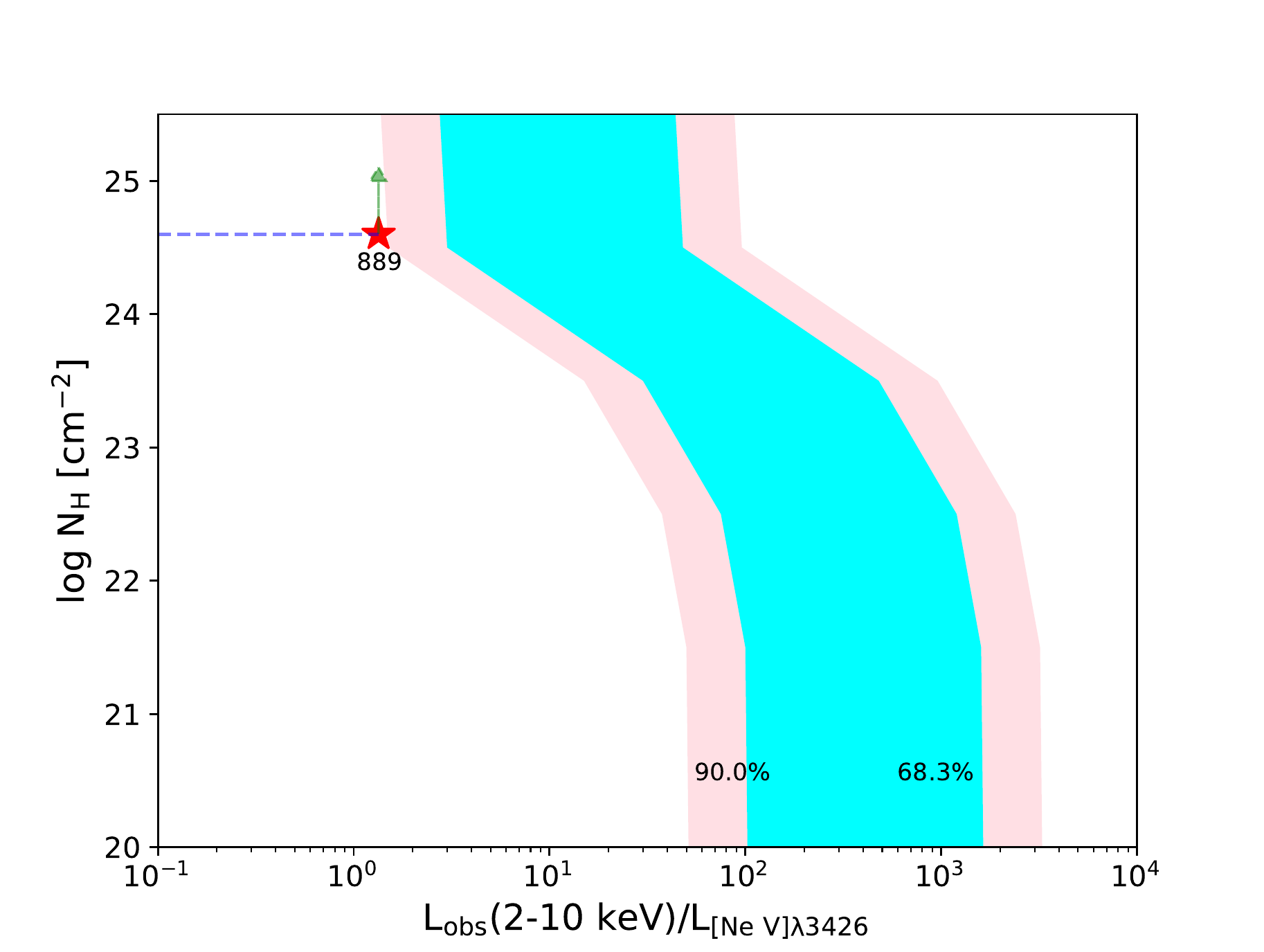}
	\caption{Left panel:  Rest-frame 2--10 keV to [O~III] luminosity ratio as a function of ${\rm N_H}$. The cyan shaded regions correspond to $68.3\%$ around the $<$X/[O~III]$>$ ratio \citep{1998A&A...338..781M}. The solid red stars are for CT AGN candidates. Right panel:  Rest-frame 2--10 keV to [Ne~V] luminosity ratio as a function of ${\rm N_H}$. The cyan and pink shaded regions correspond to $68.3\%$ and $90\%$ around the $<$X/[Ne~V]$>$ ratio \citep{2010A&A...519A..92G}.}
	\label{fig:optical-diagnostics}
\end{figure*}
\cite{1998A&A...338..781M} provided a diagnostic based on the ratio between the rest-frame 2--10 keV and the [O~III] luminosities. The left panel of Figure~\ref{fig:optical-diagnostics} shows that the ${\rm L^{obs}_{X}/L_{[O~III]}}$ ratio as a function of ${\rm N_H}$. The cyan shaded regions correspond to $\pm 1\sigma$ around the $<$X/[O~III]$>$ ratio \citep{1998A&A...338..781M}. Based on the information in the left panel of Figure~\ref{fig:optical-diagnostics}, we can derive the column densities of these two candidates. There is a 68.3\% probability that the column density of the source XID 640 is larger than ${\rm 4.0\times10^{23}\ cm^{-2}}$. Similarly, the column density of the source XID 889 is larger than ${\rm 10^{24}\ cm^{-2}}$. Besides, \cite{2010A&A...519A..92G} provided another diagnostic based on the $<$X/[Ne~V]$>$ ratio. The right panel of Figure~\ref{fig:optical-diagnostics} shows that the ${\rm L^{obs}_{X}/L_{[Ne V]}}$ ratio as a function of ${\rm N_H}$. Based on the information in the right panel of Figure~\ref{fig:optical-diagnostics}, we can derive that the column density of the source XID 889 is ${\rm N_H >4\times 10^{24}\ cm^{-2}}$ with a 90\% probability. Through the above optical diagnosis, we can identify that the source XID 889 is a CT AGN.

\subsection{Mid-IR diagnostics}\label{sec:mid-IR}

The column densities of AGNs can be estimated by comparing the X-ray to mid-IR luminosities \citep[e.g.,][]{2015A&A...578A.120L, 2015ApJ...809..115L, 2017ApJ...846...20L}.
\begin{figure}[ht]
	\includegraphics[width=1.05\linewidth]{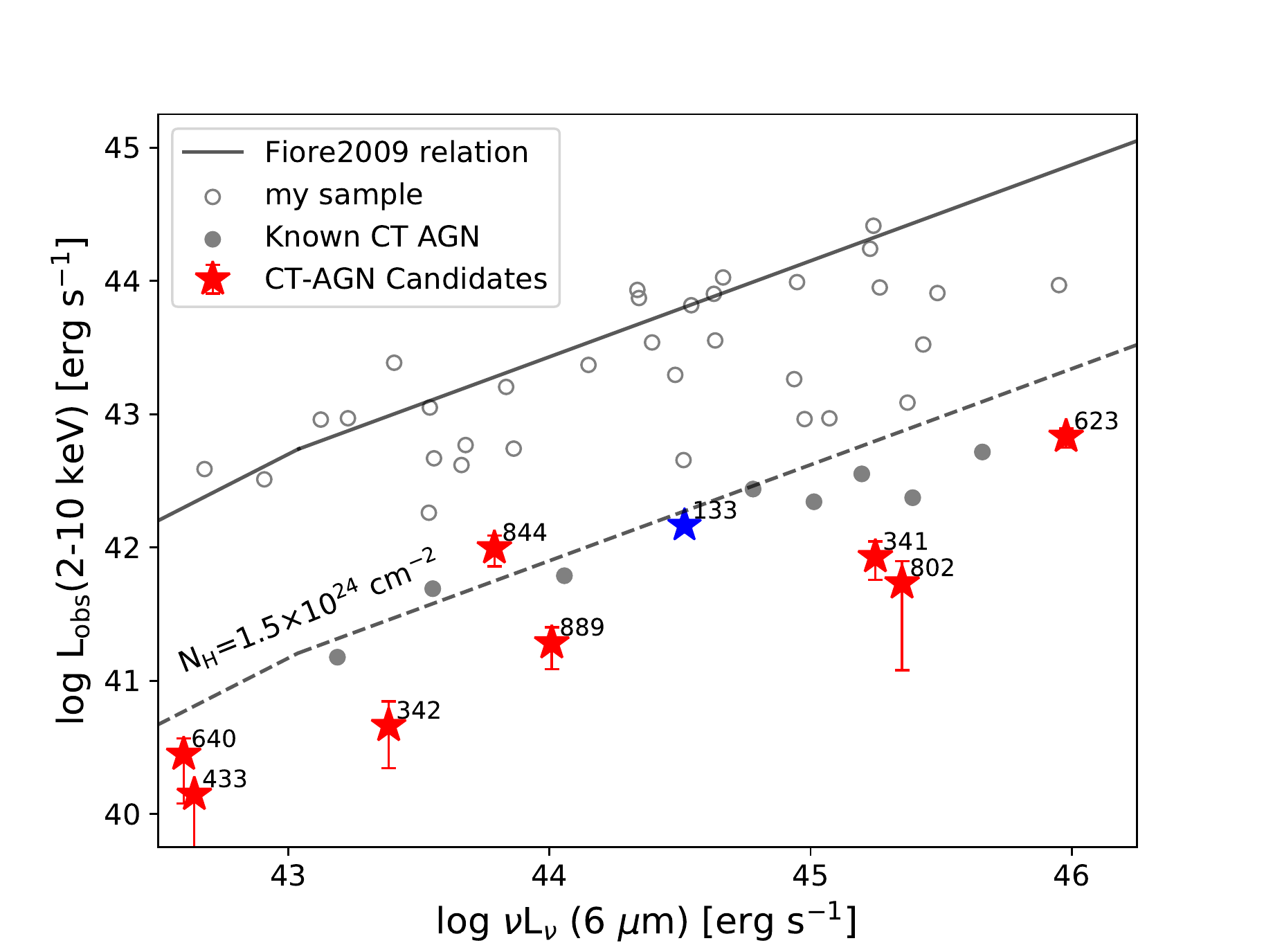}
	\caption{Rest-frame 2--10 keV observed X-ray versus rest-frame 6~\micron\ luminosity for our sample. The black line represents the relation for \cite{2009ApJ...693..447F}. The dashed line indicates the same relationship but where the X-ray luminosities are absorbed by a column density of ${\rm N_H=1.5\times 10^{24}\ cm^{-2}}$. The open grey circles are our AGN sample and the solid grey circles represent CT AGNs identified by X-ray. The solid red stars are the candidates of CT AGN. }
	\label{fig:midIR-diagnostics}
\end{figure}
Figure~\ref{fig:midIR-diagnostics} shows the distribution of observed X-ray versus 6~\micron\ luminosities for our sample. Except that the X-ray luminosities of eight candidates (solid red stars) are from Table~\ref{table:x-ray-diagno}, the X-ray luminosities of the remaining AGNs are from Table~\ref{table:sample}. The black line represents the relation from \cite{2009ApJ...693..447F}. Assuming that the intrinsic power-law spectrum has a photon index of 1.8 in the 2--10 keV energies range, the dashed line indicates the relationship after being absorbed by ${\rm N_H=1.5\times 10^{24}\ cm^{-2}}$ gas\footnote{Absorbed model : phabs.}, which means that the sources below the dashed line should be CT AGNs. 
We find that the 7 known CT AGNs are below the dashed line, indeed. 
Among eight CT AGN candidates, seven of which are below the dashed line. More interesting thing is that these seven candidates seem to have a higher column density than the eight known CT AGNs. 
One exception is XID 844 sitting above the dashed line, suggesting that this candidate may be not a CT AGN, but it is still a heavily obscured AGN. The column density of this candidate is ${\rm 1.8 ^{+0.8}_{-0.6}\times 10^{23}\ cm^{-2}}$ obtained by X-ray spectroscopy fitting \citep{2019ApJ...877....5L}, and it may be underestimated compared to the result of Figure~\ref{fig:midIR-diagnostics} (the column density of XID 844 is about ${\rm 10^{24}\ cm^{-2}}$). 
Besides, we find a source (the blue star in Figure~\ref{fig:midIR-diagnostics}, XID 133) that is not selected as a CT AGN candidate, but it is below the dashed line. 
The column density of this source is ${\rm 1.46_{-0.45}^{+0.84}\times10^{24} \ cm^{-2}}$ \citep{2019ApJ...877....5L}, its upper limit is larger than ${\rm 1.5\times10^{24}  \ cm^{-2}}$.
The column density of this source, which is estimated by comparing its X-ray to mid-IR luminosity, should be larger than ${\rm 1.5\times10^{24}  \ cm^{-2}}$. 
Combining these two cases, we can infer that the source XID 133 should be a CT AGN.

Luckily, we identify hitherto unknown 8 CT AGNs by the multi-wavelength approaches. The summary of the multi-wavelength diagnostics is listed in Table~\ref{table:new-CT-AGN}.
\begin{deluxetable}{l l c c c}[ht]
	\tablecaption{The summary of the multi-wavelength diagnostics.\label{table:new-CT-AGN}}
	\tablewidth{700pt}
	\tabletypesize{\footnotesize}
	\tablehead{
		\colhead{XID} & \colhead{z} & \multicolumn{3}{c}{Diagnostics} \\ 
		&  & \multicolumn{1}{c}{X-ray}&\multicolumn{1}{c}{Optical} & \multicolumn{1}{c}{Mid-IR} \\
		\colhead{(1)} & \colhead{(2)} & \colhead{(3)} & \colhead{(4)}  & \colhead{(5)}
	} 
	\startdata
	133 & 3.47 & & & CT  \\
	341 & 1.83 & \multicolumn{1}{c}{\nodata} & \multicolumn{1}{c}{\nodata} & CT  \\
	342 & 0.76 & \multicolumn{1}{c}{\nodata} &\multicolumn{1}{c}{\nodata} & CT  \\
	433 & 0.70 & \multicolumn{1}{c}{\nodata}&\multicolumn{1}{c}{\nodata} & CT \\
	623 & 3.88 & CT &\multicolumn{1}{c}{\nodata} & CT \\
	640 & 0.54 &\multicolumn{1}{c}{\nodata} & $>4\times 10^{23}\ {\rm cm}^{-2}$(68.3\%) & CT\\
	802 & 2.41 & \multicolumn{1}{c}{\nodata} &\multicolumn{1}{c}{\nodata} & CT \\
	844 & 1.10 & \multicolumn{1}{c}{\nodata} &\multicolumn{1}{c}{\nodata} & non-CT \\
	889 & 0.52 & \multicolumn{1}{c}{\nodata} & CT & CT \\
	\enddata
	\tablecomments{(1) and (2) are XID and redshift. (3), (4), and (5) are X-ray, optical spectrum, and mid-IR diagnostics.}
\end{deluxetable}

\section{Discussion} \label{sec:discussion}
\subsection{X-ray origin for CT AGNs}\label{sec:X-ray-origin}
For the galaxies that host AGNs, the majority of the X-ray emission originates from the center AGNs, and a small amount of X-ray emission can be contributed by the X-ray binaries (XRBs) in their host galaxies. If the central AGN is a CT AGN, then the X-ray emission from the central AGN may be completely absorbed. In order to diagnose whether the X-ray emission of 8 new CT AGNs originates from central AGNs or the XRB populations, we compare the observed X-ray and the X-ray luminosities expected from XRB populations.

The X-ray luminosity from XRB populations is composed of low-mass X-ray binaries (LMXBs) and high-mass X-ray binaries (HMXBs). Previous studies have shown that the total X-ray luminosity from the LMXB and HMXB population is proportional to M$_*$ and SFR of the host galaxy, respectively \citep[e.g.,][]{2004MNRAS.347L..57G,2016ApJ...825....7L}.  \cite{2016ApJ...825....7L} provided the following empirical relation:
\begin{equation}\nonumber
\begin{aligned}
 {\rm L_{XRB}} &= {\rm L_{LMXB} + L_{HMXB}}  \\
               &={\rm \alpha (1 + z)^\gamma M_* + \beta (1 + z)^\delta SFR},
\end{aligned}
\end{equation}
where log $\alpha = 29.30$, log $\beta = 39.40$, $\gamma = 2.19$, and $\delta = 1.02$ for 2--10 keV, respectively.

\begin{figure}[ht]
	\includegraphics[width=1.05\linewidth]{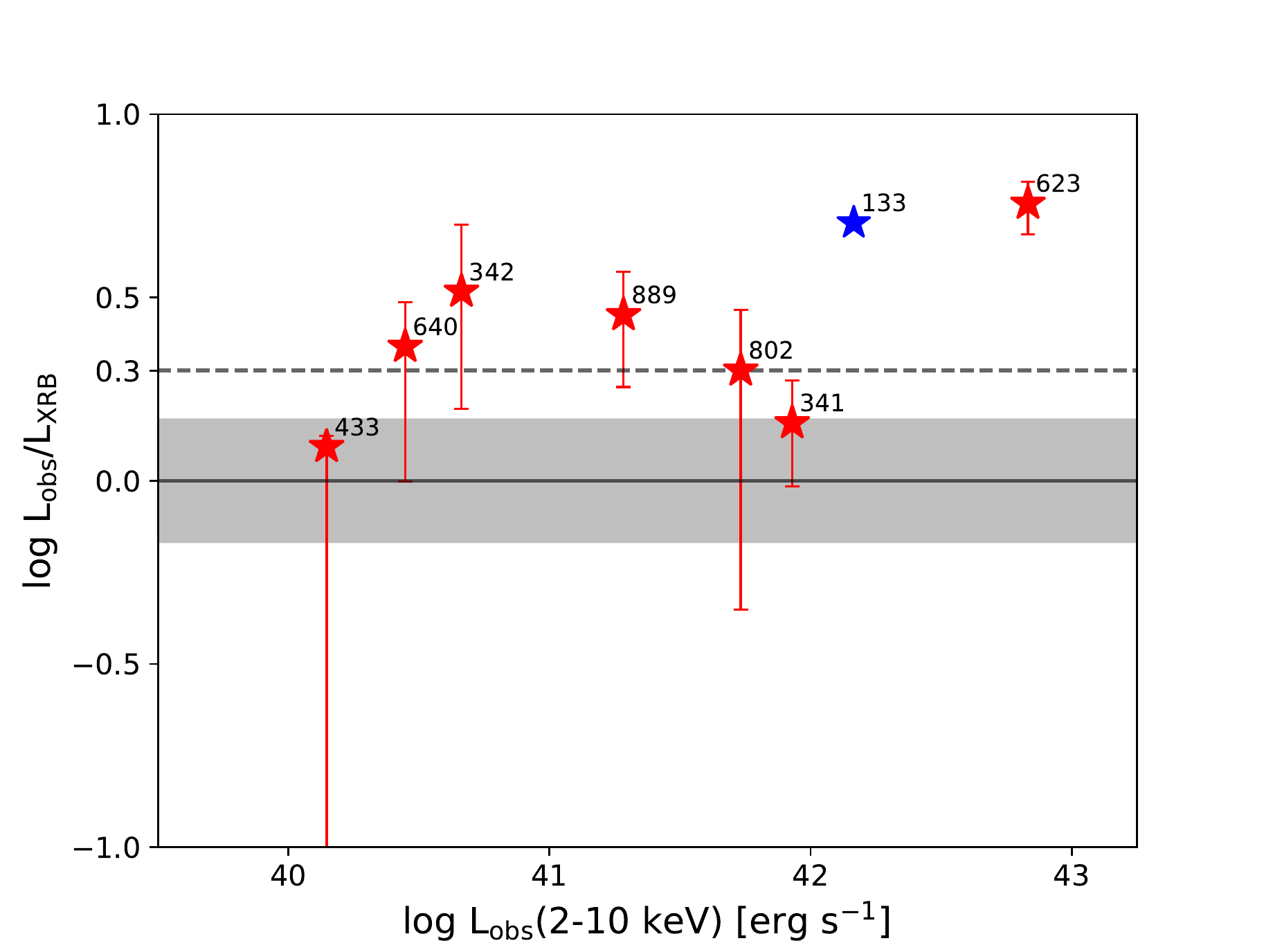}
	\caption{Observed 2--10 keV X-ray luminosity over 2--10 keV X-ray luminosity from XRB populations (${\rm L_{obs}/L_{XRB}}$) vs. observed 2--10 keV X-ray luminosity (${\rm L_{obs}}$)  for the 8 new CT AGNs. The gray shaded area shows the expected $1\sigma$ dispersion of the derived ${\rm L_{obs}/L_{XRB}}$ \citep[0.17 dex,][]{2016ApJ...825....7L}. The dashed line is ${\rm L_{obs}/L_{XRB}} = 2$. }
	\label{fig:Xray-compare}
\end{figure}
Figure~\ref{fig:Xray-compare} presents the observed X-ray over XRB X-ray luminosities as a function of observed X-ray luminosities. The gray shaded area shows the expected $1\sigma$ dispersion of the derived ${\rm L_{obs}/L_{XRB}}$ \citep[0.17 dex,][]{2016ApJ...825....7L}. 
There are two CT AGNs (XID 341 and 433) that locate in the gray shaded area, indicating that their observed X-ray emission may be entirely contributed by the XRBs in their host galaxies. 
The dashed line is ${\rm L_{obs}/L_{XRB}} = 2$, indicating that the X-ray emission from the center AGNs is equal to that from the XRBs. The CT AGNs XID 640 and 802 are near the dashed line. However, their lower limits are lower than or equal to those expected from XRB populations. We can't exclude the possibility that their X-ray emission is from the XRB populations. 
There are still four remaining CT AGNs (XID 133, 342, 623, and 889) that are above the dashed line, suggesting that the majority of their X-ray emission still originates from the center AGNs. However, the lower limits of XID 342 and 889 are below the dashed line. We cannot rule out the possibility that the emission of the XRB populations dominates the X-ray emission of these two AGNs.
Moreover, the X-ray spectrum analysis in Section~\ref{sec:X-ray-diagno} seems to assess whether the observed X-ray emssion is produced by the AGNs. The measured photon indices of four CT AGNs (XID 342, 433, 640, and 802) are $\Gamma > 3$. However, the generally measured photon index for AGN is between 1.4 and 2.6, suggesting that their observed X-ray emission should not be dominated by (or come from) the AGNs.

Combining the above information, we can infer the X-ray origin of these eight CT AGNs, and the conclusions are as follows: (1) The observed X-ray emission of two CT AGNs (XID 133 and 623) should be dominated by the center AGNs. (2) The observed X-ray emission of two CT AGNs (XID 341 and 433) may come from the XRB populations. (3) The observed X-ray emission of three CT AGNs (XID 342, 640, and 802) should not be dominated by the AGNs. (4) There is still a CT AGN (XID 889) whose X-ray emission is dominated by central AGNs with a strong possibility. However, it cannot exclude the possibility that the emission of the XRB populations dominates its X-ray emission.


\subsection{Why the $N_H$ is underestimated?}
In Section~\ref{sec:X-ray-origin}, we have demonstrated that the X-ray emission of the two CT AGNs (XID 133 and 623) mainly originates from the central AGN.
The column density of XID 133 obtained by the X-ray spectrum \citep{2019ApJ...877....5L} is reasonable within the error range (see Section~\ref{sec:mid-IR} for detailed). While the XID 623 is a high-redshift source (z = 3.88), its rest-frame energies are mainly above 10 keV. One reason may be that the X-ray spectra above 10 keV are not significantly absorbed. Moreover, the XID 623 is a radio-loud AGN, so its jet will provide a part of the X-ray emission. Previous studies \citep{2017ApJS..228....2L,2019ApJ...877....5L} did not consider the contribution of the jet, thus underestimated its column density.

The X-ray emission of two CT AGN (XID 341 and 433) might originate from the XRBs in the host galaxies and suggested that their X-ray emission originated from AGNs was entirely absorbed. Therefore, the column densities of the center AGNs cannot be obtained by fitting their X-ray spectra. Similarly, the column density of the three AGNs (XID 342, 640, and 802) cannot be also obtained, because their X-ray emission is not dominated by AGNs. However, \cite{2017ApJS..228....2L} found that the X-ray spectra of those five CT AGNs presented weak absorbed or non-absorbed. Thus the column densities of these five CT AGNs are severely underestimated.

The X-ray emission of XID 889 is uncertain whether it is dominated by the AGN. If the AGN does not dominate the X-ray emission, then the reason for its underestimation of column density should be similar to the above five CT AGNs. If the AGN dominates, its X-ray emission may be the soft scattered component of the polar region, while the X-ray emission from the nucleus is hidden. However, \cite{2017ApJS..228....2L} misunderstood this source as a weakly absorbing AGN. Thus it column density is severely underestimated.



\subsection{Selection efficiency of CT AGNs}
In the deep field survey, only about 10\% of AGNs can be identified as CT AGN by X-ray spectroscopy fitting, showing the efficiency is low.
In our sample, eight (15.6\%) CT AGNs have been confirmed by X-ray spectroscopy fitting. The seven known CT AGNs are also confirmed in this work (see Section~\ref{sec:mid-IR} for details). Besides, we identify newly 8 (15.6\%) CT AGNs by the multi-wavelength approaches. The multi-wavelength approaches of selecting CT AGNs have a higher efficiency than the X-ray spectroscopy fitting. However, the multi-wavelength approaches require that a source has multi-wavelength data and that its AGN component decomposed by the SED is reliable, so our sample is only a small part of the AGNs in the CDFS survey. In other words, the multi-wavelength approach cannot be applied to all AGNs in the CDFS survey. 
In our sample, 16 sources can be identified as CT AGN, i.e. the fraction of CT AGNs is 31.2\%, which agrees well with the theoretical expectation \citep[e.g.,][]{2015ApJ...815L..13R, 2017ApJ...846...20L}.

\subsection{Accretion of central BH of CT AGNs}
CT AGNs usually have a higher accretion rate than non-CT AGNs \citep[e.g.,][]{2010ApJ...715L..99D}. Here, we compare the Eddington ratio ${\rm \lambda_{Edd}=L_{bol}/L_{Edd}}$ of CT AGNs and non-CT AGNs in our sample. Where ${\rm L_{bol}}$ is the bolometric luminosity, ${\rm L_{Edd}}$ is Eddington luminosity which is related to BH mass (i.e., ${\rm L_{Edd}=1.3 \times 10^{38} \ M_{BH}/M_{\odot}}$ erg~s$^{-1}$).

The bolometric luminosity is the total luminosity emitted at all wavelengths by the AGN \citep[e.g.,][]{2013ApJS..206....4K,2020A&A...636A..73D}. The AGN luminosity obtained by \cite{2020MNRAS.492.1887G} is the integrated luminosity from UV to IR. The luminosity from UV to IR far exceeds the luminosity of other wavelengths.
Therefore, for Type 1 AGNs, we use the AGN luminosities instead of their bolometric luminosities. Due to Type 2 AGN emission in the UV/optical bands is known to be obscured, we estimate their bolometric luminosities by integrating the AGN emission component in the range 1--1000 \micron \ and simply rescaling the result by a factor of 1.7 \citep{2007A&A...468..603P,2018A&A...618A..28Z}.  
There are 6 AGNs with broad optical emission lines in their spectra. Therefore, we estimated the BH masses of these six AGNs with the broad emission lines \citep[H$\beta$ or Mg~II,][]{2010ApJ...724..318N,2012MNRAS.427.3081T}.
The BH masses of other AGNs are estimated to use M$_{\rm BH}$--M$_*$ relation with a scaling of 0.003 \citep{2020ApJ...889...32S}.

\begin{figure}[ht]
	\includegraphics[width=1.0\linewidth]{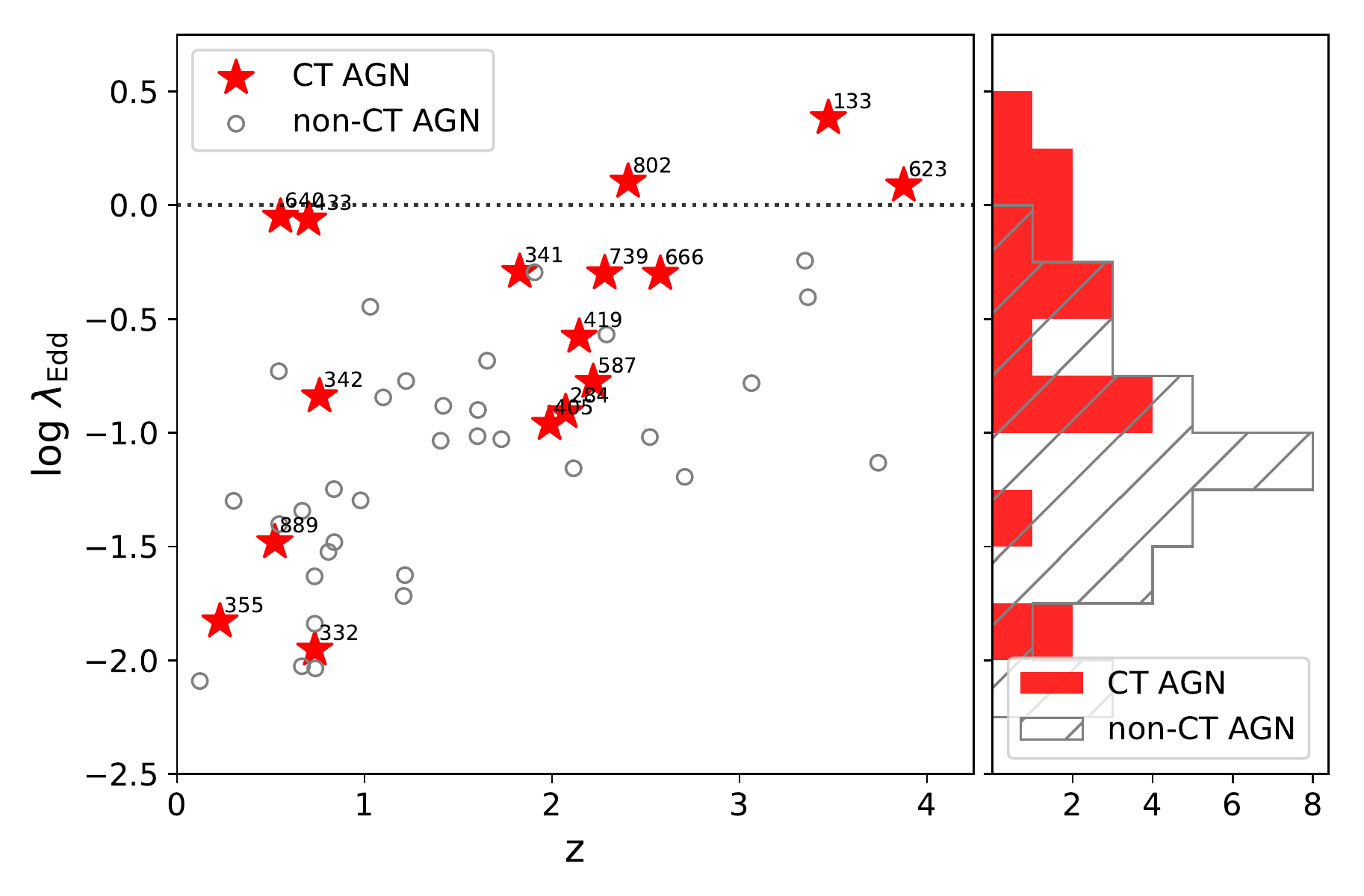}
	\caption{Left panel: Eddington ratio vs. redshift for our sample. The solid red stars indicate the 16 CT AGNs, and the open grey circles represent non-CT AGNs. The dotted line is $\lambda_{\rm Edd} = 1$. Right panel: The distribution of Eddington ratio for our sample. The red filled histogram is for the Eddington ratio of CT AGNs, and the gray histogram is for the Eddington ratio of non-CT AGNs.}
	\label{fig:eddington-ratio}
\end{figure}
The left panel of Figure~\ref{fig:eddington-ratio} presents the Eddington ratio versus redshift for our sample. Except for 3 CT AGNs (XIDs 133, 623, and 802), the Eddington ratios of the remaining AGNs are lower than 1. Interestingly, these three CT AGNs of super-Eddington accretion are all newly identified in this work. 
The multi-wavelength approaches seem to prefer the selection of CT AGN with a high Eddington ratio.
The right panel of Figure~\ref{fig:eddington-ratio} shows the distributions of the Eddington ratio for CT AGNs and non-CT AGNs. We use the Kolmogorov--Smirnov test (KS-test) to examine their distributions of Eddington ratio: the p-value is 0.010, which suggests that their distributions are different. The Eddington ratios of CT AGNs are significantly higher than those of non-CT AGNs. Our results support that the BH masses of CT AGNs are more rapid growth than those of non-CT AGNs and that CT AGNs may be a rapid growth phase of BH masses in the evolutionary scenario of AGNs \citep{2011MNRAS.411.1231G}.
\subsection{The properties of CT AGN host galaxies}
Some studies suggested that the occurrence of CT AGNs is related to the properties of their host galaxies \citep[e.g.,][]{2017MNRAS.468.1273R}. \cite{2012ApJ...755....5G} pointed out that the host galaxies of CT AGNs in the nearby universe show a high level of star formation. Therefore, we compared the SFRs and stellar masses of the host galaxies of CT AGNs and non-CT AGNs in our sample. 
\begin{figure}[ht]
	\includegraphics[width=1.05\linewidth]{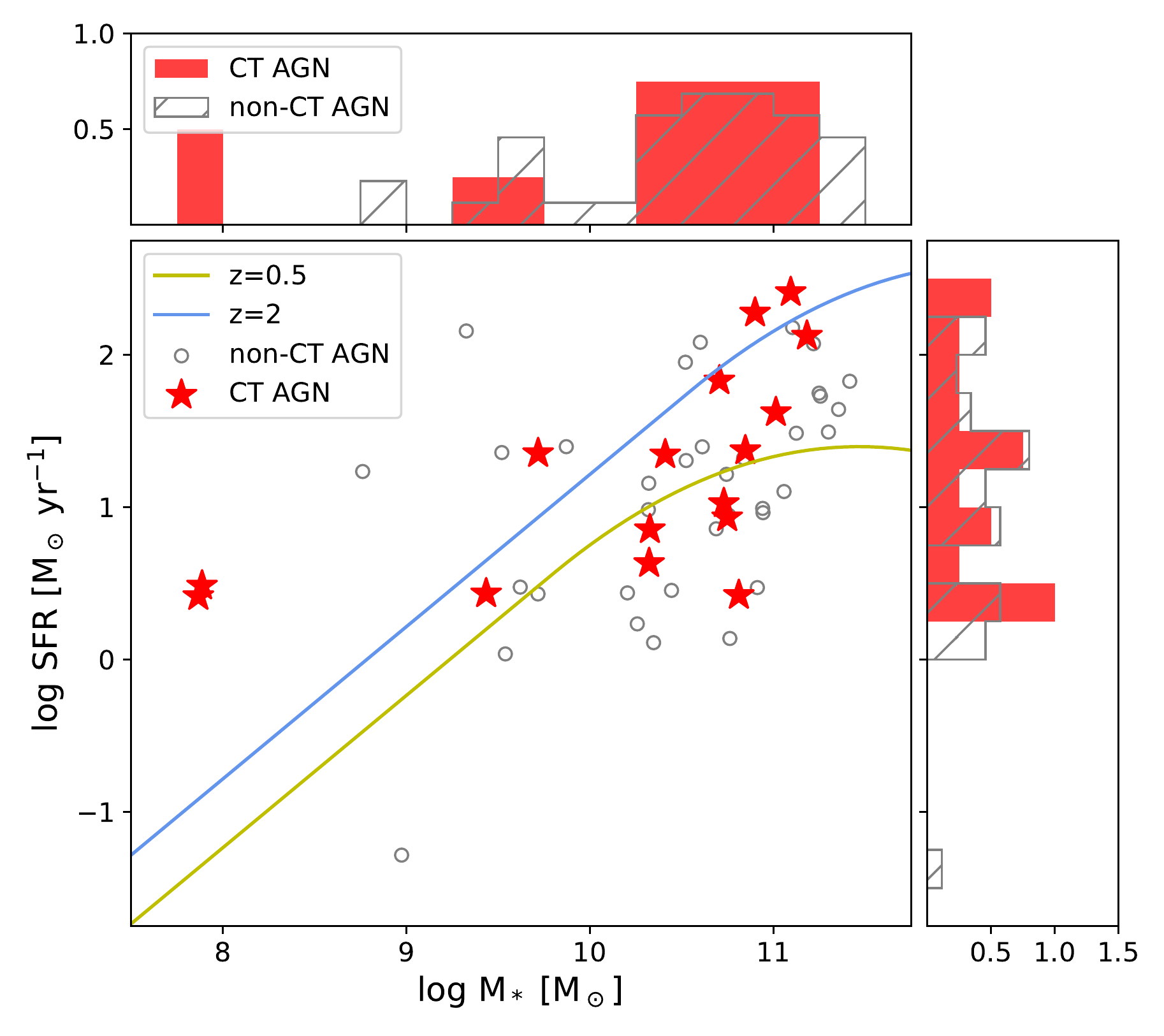}
	\caption{SFR vs. M$_*$ distribution for our sample. For comparison, the data points of 16 CT AGNs (solid red stars) and non-CT AGNs (open gray circles) are plotted. The yellow and blue lines show the main sequences of star formation at $z \sim 0.5$ and $z \sim 2$, respectively \citep{2015A&A...575A..74S}.}
	\label{fig:sfr-m}
\end{figure}
Figure~\ref{fig:sfr-m} presents the relation between SFRs and stellar masses for the host galaxies. 
We also use the KS-test to examine their stellar mass distributions, and the p-value is 0.87.  We repeat the KS-test to examine their SFR distributions, and the p-value is 0.97. These suggest that the properties of CT AGN host galaxies are similar to those of non-CT AGN host galaxies in our sample. 
Our results are against the claim of \cite{2012ApJ...755....5G}, and the main reason may be that most of the sources in our sample are high redshift AGNs. The environment of  AGN host galaxies at high-redshift may be different from the local universe.
In addition, our sample contains fewer CT AGNs, we cannot rule out our results are biased in sample selection.
In subsequent work, we will expand the CT AGN sample and further study whether there are differences in the host galaxies of both CT AGN and non-CT AGN.

\section{Summary}\label{sec:summary}
In this work, we find out missed CT AGNs from the CDFS survey using the multi-wavelength techniques and discussed their properties.
Firstly, we construct a sample containing 51 AGNs with abundant multi-wavelength data. 
We select hitherto unknown 8 CT AGN candidates using the method of the mid-IR excess. 
Diagnosis based on the characteristics of their X-ray spectra,  only the X-ray spectrum of source XID 623 shows all characteristics of CT AGNs.
Through the optical spectrum diagnosis, the source XID 889 can be identified as CT AGN, while the source XID 640 can only obtain a lower limit of the column density. Except for the source XID 844, the remaining seven candidates can be identified as CT AGN in the mid-infrared diagnosis. Besides, a new CT AGN (XID 133) is also found in section~\ref{sec:mid-IR}. 
Subsequently, we discuss the X-ray origin of these eight CT AGNs. 
Except that the X-ray emission of the two sources (XID 133 and 623) is still from the central AGNs, most or all of the X-ray emission of the remaining six sources originates from the XRB population in their host galaxies.
We also discuss the reason why their column densities were underestimated in previous studies.
We find that the multi-wavelength approaches of selecting CT AGNs are highly efficient, provided the high quality of observation data. Finally, we compare CT AGNs and non-CT AGNs in our sample. We find that CT AGNs have a higher Eddington ratio than non-CT AGNs, and that both CT AGNs and non-CT AGNs show similar properties of host galaxies.

\acknowledgments

We sincerely thank the anonymous referee for useful suggestions. We thank Bin Luo for helpful discussions. 
We acknowledge financial support from the National Key R\&D Program of China grant 2017YFA0402703 (Q.S.Gu) and National Natural Science Foundation of China grant 11733002 (Q.S.Gu). 
This research makes use of data from \textit{Chandra} Deep Field-South Survey.
We acknowledge the extensive use of the following Python packages:
\\
\textit{Software}: numpy, pandas \citep{mckinney-proc-scipy-2010},  matplotlib \citep{2007CSE.....9...90H}, astropy\citep{2013A&A...558A..33A}, Code Investigating GALaxy Evolution \citep{2019A&A...622A.103B}.

\bibliography{CT-AGN}{}
\bibliographystyle{aasjournal}



\end{document}